\documentclass[iop,apj]{emulateapj}

\slugcomment{Submitted to the Astrophysical Journal}

\shorttitle{Water fountain IRAS 15103$-$5754}
\shortauthors{G\'omez et al.}

\begin{document}


\title{The first ``water fountain'' collimated outflow in a planetary nebula}


\author{Jos\'e F. G\'omez\altaffilmark{1}, Olga Su\'arez\altaffilmark{2}, Philippe Bendjoya\altaffilmark{2}, J. Ricardo Rizzo\altaffilmark{3}, Luis F. Miranda\altaffilmark{1,4},  James A. Green\altaffilmark{5,6}, Lucero Uscanga\altaffilmark{7},  Enrique Garc\'{\i}a-Garc\'{\i}a\altaffilmark{3}, Eric Lagadec\altaffilmark{2}, Mart\'{\i}n A. Guerrero\altaffilmark{1}, Gerardo Ramos-Larios\altaffilmark{8} }


\altaffiltext{1}{Instituto de Astrof\'{\i}sica de Andaluc\'{\i}a, CSIC, Glorieta de la Astronom\'{\i}a s/n, E-18008, Granada, Spain}
\altaffiltext{2}{Laboratoire Lagrange, UMR 7293, Universit\'e de Nice Sophia-Antipolis, CNRS, Observatoire de la C\^ote d'Azur, 06304, Nice, France}
\altaffiltext{3}{Centro de Astrobiolog\'{\i}a (INTA-CSIC),
Ctra. M-108, km.~4, E-28850 Torrej\'on de Ardoz, Spain}
\altaffiltext{4}{Universidad de Vigo, Departamento de F\'{\i}sica aplicada,
   Facultad de Ciencias, Campus Lagoas-Marcosende s/n, 36310, Vigo,
   Spain}
\altaffiltext{5}{CSIRO Astronomy and Space
Science, Australia Telescope National Facility, PO Box 76, Epping, NSW
1710, Australia}
\altaffiltext{6}{SKA Organisation, Jodrell Bank Observatory, Lower Withington, Macclesfield SK11 9DL, UK}
\altaffiltext{7}{Institute of Astronomy, Astrophysics, Space Applications and Remote Sensing, National Observatory of Athens, 15236 Athens, Greece}
\altaffiltext{8}{Instituto de Astronom\'{\i}a y Meteorolog\'{\i}a, Av. Vallarta No. 2602, Col. Arcos 
Vallarta, C.P. 44130 Guadalajara, Jalisco, Mexico}


\begin{abstract}
``Water fountains'' (WFs) are evolved objects  showing high-velocity, collimated jets traced by water maser emission. Most of them are in the post-Asymptotic Giant Branch and they may represent one of the first manifestations of collimated mass loss in evolved stars.
We present water maser, carbon monoxide, and mid-infrared spectroscopic data (obtained with the Australia Telescope Compact Array, Herschel Space Observatory, and the Very Large Telescope, respectively) toward  IRAS 15103$-$5754, a possible planetary nebula (PN) with WF characteristics. Carbon monoxide observations show that IRAS 15103$-$5754 is an evolved object, while the mid-IR spectrum displays unambiguous [Ne{\sc ii}] emission, indicating that photoionization has started and thus, its nature as a PN is confirmed. Water maser spectra show several components spreading over a large velocity range ($\simeq 75$ km s$^{-1}$) and tracing a collimated jet. This indicates that the object is a WF, the first WF known that has already entered the PN phase. However, the spatial and kinematical distribution of the maser emission in this object  are significantly different from those in other WFs. Moreover, the velocity distribution of the maser emission shows a  ``Hubble-like'' flow (higher velocities at larger distances from the central star), consistent with a short-lived, explosive mass-loss event. This velocity pattern is not seen in other WFs (presumably in earlier evolutionary stages). We therefore suggest that we are witnessing a fundamental change of mass-loss processes in WFs, with water masers being pumped by steady jets in post-AGB stars, but tracing explosive/ballistic events as the object enters the PN phase.
\end{abstract}





\section{Introduction}

``Water fountain'' (WF) stars are low- to intermediate-mass stars ($< 8$ M$_\odot$) evolved objects that show collimated
jets traced by high-velocity water masers at 22 GHz \citep[see][for reviews]{ima07,des12}. The velocity spread in
their water maser spectra is typically $\ge 75$ km s$^{-1}$, and can
be as large as $\simeq 500$ km s$^{-1}$ \citep{gom11}.  These velocities indicate the presence of motions faster than the typical expansion velocities in spherical AGB envelopes, of $\simeq 10-20$ km s$^{-1}$, as traced by double-peaked OH maser spectra \citep[with velocity separations between both components of $\simeq 20-40$  km s$^{-1}$,][]{tel89}.
 Most WFs are
in the short transition phase between the Asymptotic Giant Branch
(AGB) and the planetary nebula (PN) stage. Their being in this particular
evolutionary stage, together with the short dynamical ages of the
maser outflows \citep[5-100 yr;][]{ima07}, and their relatively strong optical
obscuration \citep{sua08} indicate that WFs may represent one of the first
manifestations of collimated mass-loss in evolved stars of low and intermediate mass.

There is growing consensus that the action of collimated jets on the circumstellar envelope during the post-AGB phase will determine the shape of PNe once photoionization starts \citep{sah98,buj01}.
Therefore, WFs are key objects to understand when and how the
spherical symmetry seen in AGB stars and earlier stages is broken to
give rise to the spectacular variety of morphologies displayed in the
PN phase.

Only 16 sources have been identified so far as possible WFs \citep{des12,vle14}.
 Fourteen of them seem to be in the post-AGB phase. Of the remaining two candidates, the source W43A may be at the late AGB stage \citep{ima02}. At the other end of the age spectrum, \cite{sua09} detected a WF candidate toward the source IRAS 15103$-$5754 (hereafter I15103), which had been classified as a possible PN \citep{van93}. If confirmed, this would be the first known PN with WF characteristics. This confirmation is important, since it would prove that collimated jets with enough energy to pump maser emission can persist after the star enters the PN phase. Such a jet could contribute to the shaping of the nebula while photoionization already proceeds. 
 
I15103 is an optically obscured object \citep{sua06,rl12} that shows a bipolar morphology in near and mid-infrared images \citep{rl12,lag11}. Its classification as a PN was based on the presence of radio continuum emission \citep{van93}, which is typically seen in the photoionized regions of these objects. However, the radio continuum flux density decreases at higher frequencies \citep{urq07a,sua14}, which is indicative of non-thermal emission and cannot be attributed to free-free processes in a photoionized region. The time-evolution of its non-thermal emission has been attributed to the onset of photoionization \citep{sua14}.
 However, to date there is no conclusive evidence that I15103 has already entered the PN phase or even that it is an evolved object.

In this paper, we provide solid evidence that I15103 is a bona fide WF, and that it is also PN. Our observations are presented in Sec. \ref{sec_obs}. In Sec. \ref{sec_discussion} we present and discuss our results, and we summarize our conclusions in Sec. \ref{sec_conclusions}

\section{Observations}

\label{sec_obs}

\subsection{Australia Telescope Compact Array}

\label{sec_atca}
Observations at 1.3 cm were carried out with the Australia Telescope Compact Array\footnote{The Australia Telescope is funded by the Commonwealth of Australia for operation as a National Facility managed by CSIRO} (ATCA) on 2011 August 11 and 2012 October 23, under observing projects C2315 and C2746, respectively. We used the 
Compact Array Broadband Backend (CABB), which provides two independent
bands of 2 GHz each, in dual linear polarization. We centered these bands at 22 and 24 GHz. The observations on
2011 were carried out in the H168 array configuration (including antenna C06, located at 6 km from the array center), and used the 64M mode of CABB, which samples each 2 GHz
bandpass into 32 frequency channels of 64 MHz. Observations on 2012 October were carried out in the H214 array configuration (without antenna C06) and
used the 1M mode, which samples 2048 channels, each of them 1 MHz
wide. In addition to this broad band mode \citep[whose data were presented in][]{sua14}, we observed the
emission of the $6_{16}-5_{23}$ transition of H$_2$O
(rest frequency = 22235.08 MHz), using the zooming capabilities of CABB: on 2011 August, one of
the wideband channels (64 MHz width) was subdivided into 2048
spectral-line channels, providing a velocity resolution of
$\simeq 0.42$ km s$^{-1}$, and a total coverage of $\simeq 860$ km
s$^{-1}$, while on 2012 October, we combined 16 wideband channels (1 MHz width), spaced by 0.5 MHz, and each subdivided into 2048 spectral-line channels, thus providing a velocity
resolution of $\simeq 6.6\times 10^{-3}$ km s$^{-1}$, and a total coverage of
$\simeq 114$ km s$^{-1}$. For the latter epoch of observations, spectral line data presented in this paper were smoothed up to a velocity resolution of $\simeq 0.066$ km s$^{-1}$ in our data reduction.

The absolute flux, bandpass, and complex gains were calibrated with sources PKS 1934-638, PKS 1921-293, and PKS
1613-586, respectively.
Initial calibration and imaging
was carried out with standard procedures using MIRIAD \citep{sau04}. Continuum was subtracted from line data with task UVLIN of MIRIAD. Then, we
selected the spectral-line channel with the highest flux density, and applied
self-calibration with the AIPS package of NRAO. The derived amplitude and phase
corrections were then applied to the whole line dataset. A new
narrow-band continuum dataset at $\simeq 22$ GHz was obtained by averaging the
line-free channels from this self-calibrated dataset. Since these narrow-band continuum data share the same corrections as obtained for the self-calibration of
the maser line, the systematic astrometric errors and those
due to atmospheric phase variations are the same in line and
continuum. In this way, we can ascertain the relative position between
the maser components and the maximum of the continuum emission with an
accuracy which is only limited by the noise in the images. Under these
conditions, the 1-sigma positional accuracy between the continuum and
the maser components (and among the different maser components) can be
quantified as $\sigma_{\rm pos} \simeq \theta /(2\times \rm snr)$,
where $\theta$ is the source size in the image (i.e., the synthesized
beam in the case of point sources), and snr is the signal-to-noise
ratio. Considering maser components $> 1$ Jy, these accuracies were typically better than
$\simeq 30$ and 25 milliarcsec between maser and continuum for our data on 2011 and 2012,
respectively, and $\simeq 1$ and 20 milliarcsec among maser components, for 2011 and 2012 data, respectively.

Final images of line and continuum emission were obtained with a robust weighting of visibilities, using a robust parameter 0 in task IMAGR of AIPS. Synthesized beam sizes were $\simeq 3.1''\times 1.1''$ (position angle = $0^\circ$) and $10.0''\times 7.0''$ (position angle = $64^\circ$) in 2011 and 2012, respectively.

\subsection{Herschel Space Observatory}

\label{sec_herschel}

I15103 was included as one of the targets of {\sl Herschel}
Space Observatory\footnote{{\it Herschel} is an ESA space observatory with science instruments provided by European-led Principal Investigator consortia and with important participation from NASA.} \citep{Pil10} in the first open time period (Proposal ID: OT1\_jrizzo\_1). The CO observations were obtained using
the Heterodyne Instrument for the Far-Infrared (HIFI)  \citep{Gra10}. We present here the first analysis of
the CO $J=5\rightarrow4$ (567.2684~GHz, band 1) and $10\rightarrow9$
(1151.985~GHz, band 5) spectra. Full width at half maximum (FWHM) of the
telescope beam is estimated to be $44''$ and $19''$ at these bands, respectively.

The observations were gathered in two orthogonal lineal polarizations.
The receiver worked in double side-band (DBS) mode, which allowed
the use of both lower- and upper-side bands simultaneously. The
observing mode was dual beam switching, with the emission-free position
 $3'$ away from the source position. We used the HIFI wide-band spectrometer
(WBS), which provides four adjacent bands per polarization
with an instantaneous width of 1~GHz each.

The spectra were initially processed using HIPE ({\sl
Herschel} Interactive Processing Environment), version 11.0.1, and
exported to CLASS\footnote{GILDAS software.
http://www.iram.fr/IRAMFR/GILDAS/} using the hiClass tool within HIPE.
Each 1-GHz individual band was concatenated providing 2 individual
spectra (one per polarization) of 4~GHz total bandwidth. The original frequency spacing was
500~kHz, although the effective spectral resolution was $\simeq 1.1$~MHz.
A baseline of low order was fitted, and there was no evidence of
significant standing waves. Spectra have been converted to a main-beam
brightness temperature scale ($T_\mathrm{mb}$), by assuming main beam efficiencies
of 0.76 and 0.64 for bands 1 and 5,
respectively \citep{Roe12}.

After spectral Hanning smoothing, the resulting  spectra we present in this paper have $1\sigma$ rms noises of 55~mK and 52~mK, and velocity resolutions of 3.1 km s$^{-1}$ and 
4.7 km s$^{-1}$ for the CO $J=5\rightarrow 4$ and $10\rightarrow 9$ transitions, respectively.

\subsection{Very Large Telescope}

A mid-infrared, N-band (8-13 $\mu$m) spectrum with spectral resolution of 300 was taken toward I15103 on 2009 June 24 with the VLT spectrometer and imager for the mid--infrared (VISIR) instrument \citep{lag04}, mounted on the Very Large Telescope (VLT) UT3.  The slit (with a length of $32''$ and width of $0.4''$) was centered on the object and aligned along the major axis of the nebula, with a position angle of 45$^\circ$. 
The classical chopping and nodding technique was used to remove the sky background, using a parallel chopping with a frequency of 0.25 Hz and a ``chop throw'' of $10''$. To cover the full N band window, we used four gratings centered at 8.5, 9.8, 11.4, and 12.2 $\mu$m. We used detector integration times of 0.2 seconds and a total integration time of 180 seconds per grating. In order to remove telluric lines, we observed the mid-infrared standard star HD 111915 with exactly same settings immediately after the  I15103 observations.
Data reduction was performed using Gasgano  \citep[v2.4.5;][]{izz04}. 

\section{Results and discussion}

\label{sec_discussion}

\subsection{Radio continuum emission}

\label{sec_continuum}
Both broad-band \citep{sua14} and narrow-band continuum (this paper)
images show an unresolved source of emission at 22 GHz, located at
R.A.(J2000) $= 15^h 14^m 17.94^s$, Dec(J2000) $=-58^\circ 05' 21.9''$
(absolute positional error $\simeq 0.6''$), with a flux density of $\simeq 35\pm 5$ mJy. 
The narrow-band continuum
images were used to determine the relative position of the maser
components with respect to the continuum emission, with a high
accuracy (as mentioned in Section \ref{sec_atca}). 
\cite{sua14} have
confirmed that the radio continuum emission is non-thermal, and its
spectral index has significantly varied over the past few years. The
non-thermal emission traces the motion of electrons, accelerated by
shocks, within a magnetic field, and the variation of spectral index
has been interpreted by \cite{sua14} as the result of the onset of
photoionization.



The highest angular resolution in our observations is of $\simeq 3''\times 1''$, which is larger than the extent of the infrared lobes of the nebula 
\citep{lag11},
so we cannot accurately determine the exact location and extent of this non-thermal emission. However, assuming that mass-loss processes are symmetrical with respect to the central star, the position of the unresolved radio continuum emission provides an estimate of the location of the star. 

\subsection{The outflow traced by water masers}

\label{sec_maser}

\begin{figure*}
\plotone{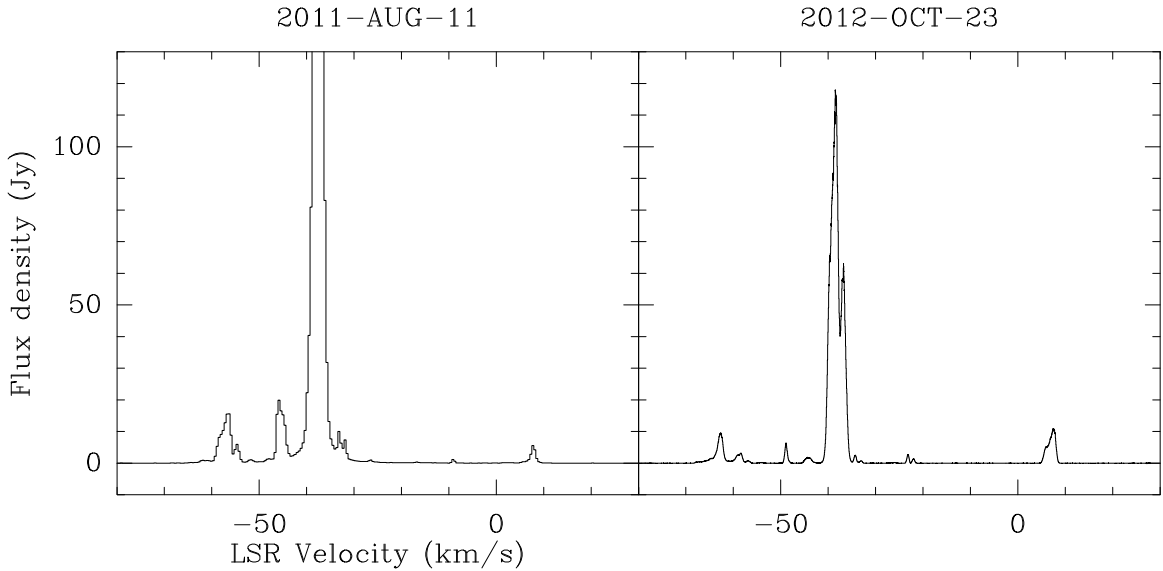}
\caption{Water maser spectra toward I15103, in two different epochs, integrating over the whole emitting region. The component with the highest flux density (1.731 kJy) in the left panel has been truncated, so that weaker components can be seen. \label{fig_spectra}}
\end{figure*}

\begin{figure*}
\plottwo{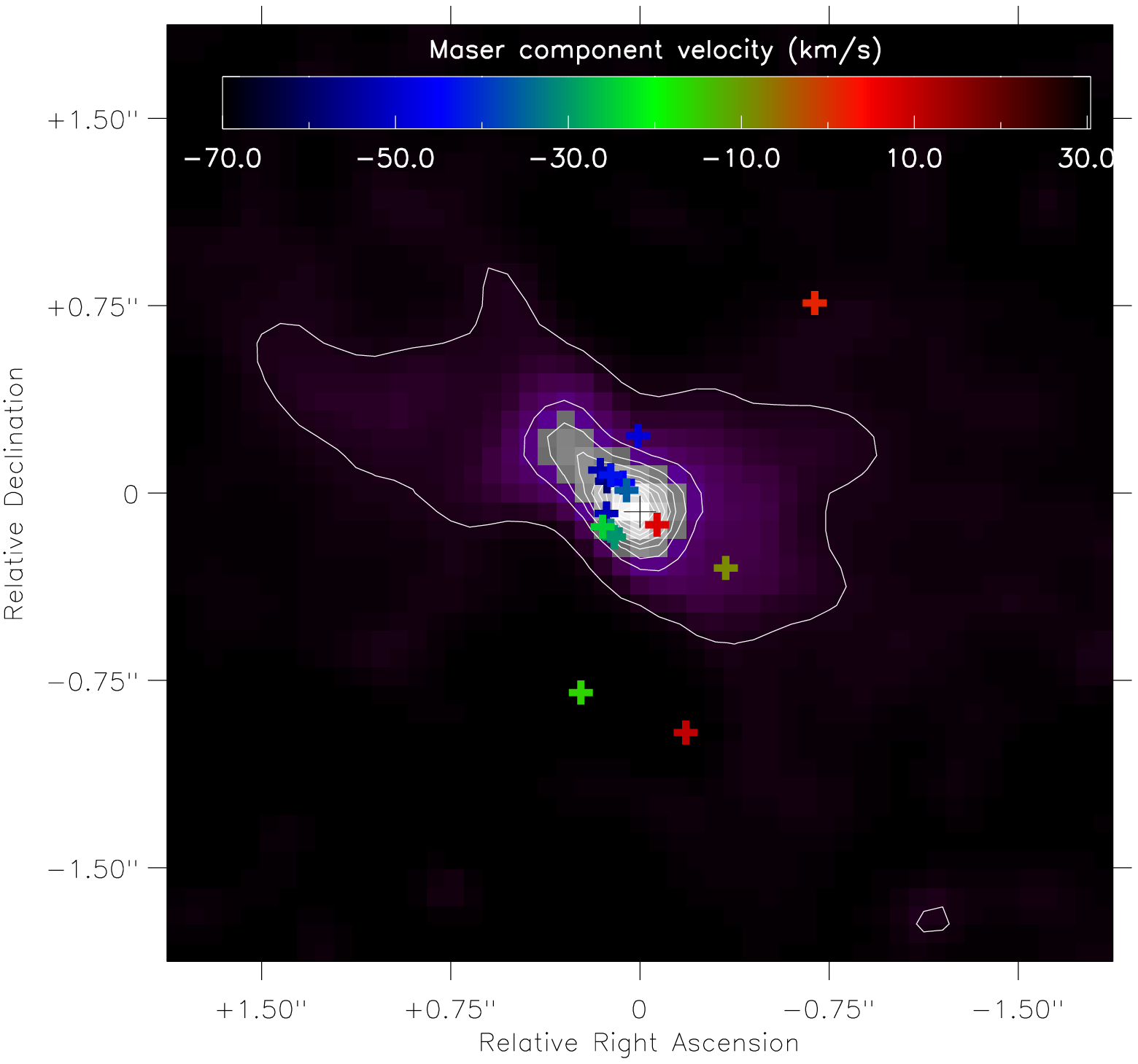}{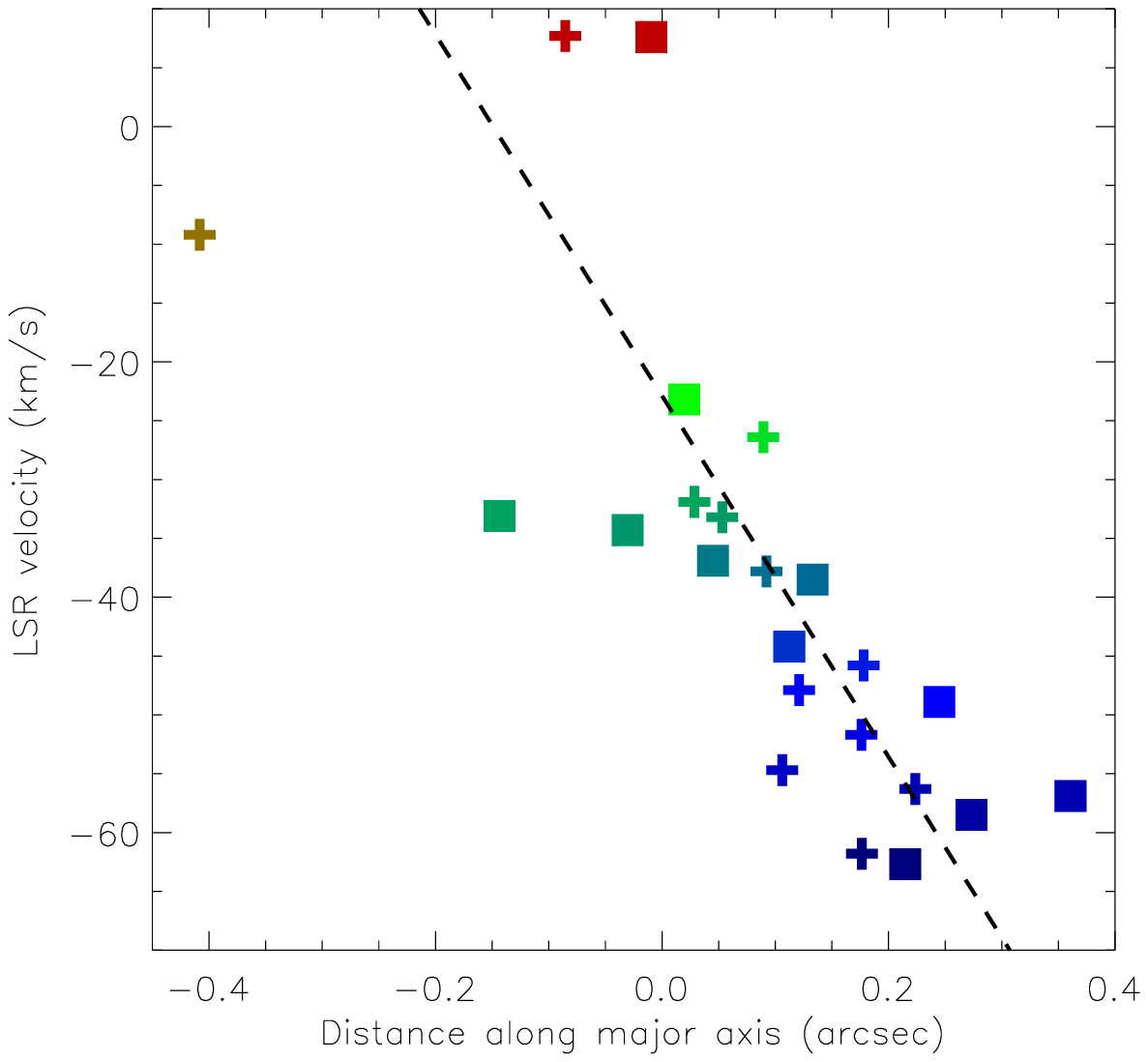}
\caption{Left: Water maser components (colored crosses) in the 2011 August data, plotted on an image the 12.81 $\mu$m ([Ne {\sc ii}] filter) presented by \cite{lag11}. The color scale of the crosses represent the LSR velocity of the components. Coordinates are offsets with respect to the position of the radio continuum emission (marked with a black cross).   The alignment between the radio data and the mid-IR image has been made assuming that the intensity peak of the latter coincides with the position of the
  radio continuum emission. Right: LSR velocity of the maser components vs. projected distance to the radio continuum emission, along the major axis of the maser structure (P.A. $=56^\circ$). Crosses and closed squares represent the data on 2011 and 2012, respectively. Color code is the same as in the right panel. The dashed line represents the
 best linear fit to the plotted data, with distances weighted by the inverse squared positional uncertainty of each point. \label{fig_image}}
\end{figure*}

Maser emission was detected in both epochs, with multiple components spanning over a velocity range of up to $\simeq 75$ km s$^{-1}$ (Fig. \ref{fig_spectra} and Table \ref{tab_maser}). In 2011 the peak emission was extraordinarily intense, reaching 1.731 kJy. The data on 2011 (which provides the highest positional accuracy) show that most of the maser components are aligned with the radio continuum position on a linear distribution (Fig. \ref{fig_image}), along a position angle of $\simeq 56^\circ$ (measured from north to east). Four additional components show a parallel linear distribution, offset $\simeq 0.14''$ south east of the radio continuum position. Four maser components depart from these main ``strings'', and they could be part of a different structure (like an equatorial torus), but they are the weakest ones and, hence, their positional accuracy is lower. Therefore, a significant fraction of the water maser emission seems to trace a high-velocity, collimated ejection, which confirms its nature as a water fountain.  The direction of
the water maser linear distribution
 is almost coincident with that of the lobes detected in the IR
images \citep[see Fig. \ref{fig_image}]{lag11} and presents an extension of $0.4''$
from the center of the PN, equivalent to 440 AU assuming a
kinematical distance of 1.1 kpc \citep[near kinematical distance for a $V_{\rm LSR}\simeq -23$ km s$^{-1}$, see below, and assuming the model by][without applying the correction for high-mass star-forming regions]{rei09}. The high velocities of the maser components (total velocity spread $\simeq 75$ km s$^{-1}$), together
with the small spatial extent of the structure they trace, suggest that the collimated ejection is
extremely young. However, with the information on LSR velocity and projected size alone it is impossible to obtain an accurate estimate of the dynamical age of the outflow. For an age estimate, we would need proper motion measurements, or an accurate determination of the distance and the inclination angle  with respect to the observer.

{
\begin{deluxetable}{rrrrr}
\tablecolumns{5}
\tablewidth{0pc}
\tablecaption{Water maser components in IRAS 15103$-$5754\label{tab_maser}}
\tablehead{
\colhead{$V_{\rm LSR}$}    &  \colhead{$S_\nu$} &   \colhead{$\Delta_{ra}$}   
& \colhead{$\Delta_{dec}$}  & \colhead{$\sigma_p$}\\
\colhead{(km s$^{-1}$)}     & \colhead{(Jy)}    &   \colhead{(mas)} & \colhead{(mas)} & \colhead{(mas)}
}
\startdata
\multicolumn{5}{c}{2011 August 11}\\
\cline{1-5}
$-61.8$  & $0.890\pm 0.007$ & 131 & 122 & 4\\	
$-56.8$  & $14.200 \pm 0.009$ & 157.0 & 167.2 & 0.3\\
$-54.7$  & $3.918 \pm 0.009$ & 132.6 & $-6.8$ & 1.1\\
$-51.7$  & $0.695 \pm 0.007$ & 7 & 304 & 5\\
$-47.9$  & $1.422 \pm 0.007$ & 67.0 & 117.2 & 2.2 \\
$-45.8$  & $11.847 \pm 0.010$ & 115.7 & 147.0 & 0.4\\
$-37.8$  & $1731.1 \pm 0.3$\phn\phn & 53.20 & 86.00 & 0.09\\
$-33.2$  & $7.823 \pm 0.010$ & 115.6 & $-76.4$ & 0.6\\
$-31.9$  & $4.885 \pm 0.008$ & 102.3 & $-101.0$ & 0.8 \\
$-26.4$  & $0.622 \pm 0.007$ & 148 & $-60$ & 5\\
$-16.7$  & $0.287 \pm 0.007$ & 234 & $-724$ & 11\\
$ -9.2$  & $0.589 \pm 0.007$ & $-340$ & $-225$ & 6\\
$  1.4$  & $0.133 \pm  0.006$ & $-692$ & 836 & 22\\
$  7.7$  & $3.580 \pm 0.007$ & $-67.8$ & $-52.2$ & 1.0\\
$ 11.1$  & $0.088 \pm  0.006$ & $-182$ & $-884$ & 30\\
\cutinhead{2012 October 23}
$-62.66$ & $9.708\pm 0.016$ & 188 & 108 & 4\\
$-58.47$ & $3.107\pm 0.014$ & 192 & 198 & 12\\
$-56.88$ & $0.802\pm 0.009$ & 221 & 303 & 30\\
$-54.00$ & $0.231\pm 0.006$ & 101 & $-577$ & 70\\
$-48.88$ & $6.120\pm 0.016$ & 124 & 239 & 7\\
$-44.31$ & $1.714\pm 0.011$ & 166 & $-30$ & 17\\
$-38.49$ & $115.53\pm 0.09$\phn & 114.0 & 70.0 & 2.0\\
$-36.75$ & $61.32\pm 0.05$\phn & 61.0 & $-5.0$ & 2.1\\
$-34.23$ & $2.468\pm 0.012$ & 31 & $-89$ & 12 \\
$-33.13$ & $0.762\pm 0.008$ & $-161$ & $-27$ & 26\\
$-26.24$ & $0.221\pm 0.010$ & 219 & $-565$ & 200 \\
$-23.13$ & $2.679\pm 0.008$ & 0 & 32 & 7\\
$-22.04$ & $1.312\pm 0.012$ & 518 & 47 & 23\\
  $7.64$ & $10.725\pm 0.023$ & $-70$ & 73 & 6\\
 \enddata
\tablecomments{Column 1 is the LSR velocity of the component peak emission. Column 2 is the flux density of the component peak emission, with its $2\sigma$ uncertainty due to the rms of the map. This provides a relative flux uncertainty between components, whereas the absolute flux calibration uncertainty is $\simeq 10$\%. Columns 3 and 4 represent the positional offsets (in right ascension and declination, respectively) of each maser component with respect to the position of the radio continuum emission. Column 5 is the $1\sigma$ relative positional uncertainty of each maser component. The actual relative accuracy between a pair of components should be obtained by adding quadratically their corresponding values in this column.}
\end{deluxetable}
} 

In Fig. \ref{fig_image} we can see that the maser emission shows
a Hubble-flow pattern, i.e., a linear dependence of velocity on distance from the radio continuum position, with increasing relative velocities at larger
distances. This linear trend gives us
an estimate of $\simeq -23$ km s$^{-1}$ for the (LSR) radial velocity of the central star (the intersect velocity of the linear fit at spatial offset = 0),
and indicates that most components are blueshifted with respect to
it. This blueshifted bias might represent an intrinsic asymmetry in the ejection, but it is more likely to be an effect of maser amplification, since the approaching side  would have the radio continuum in the background along the line of sight. Maser emission will then be favored on that side, since it amplifies a stronger background than in the redshifted side. A blueshifted bias is also seen in OH maser emission from PNe \citep{usc12} and high-mass young stellar objects (YSOs) \citep{cas11}, and it may have the same physical origin.

The observed kinematical pattern (the Hubble-like flow) is common in spectroscopic data at optical/infrared wavelengths \citep[e.g.,][]{cor93,mea05,mea08,hri08} and CO molecular lines \citep{buj97,alc01} of bipolar PNe and post-AGB stars. However, 
it is not clearly present in water maser jets of post-AGB WFs \citep{ima02,ima04,bob05,gom11}. 
Several models have been proposed to explain these linear velocity gradients in PNe and post-AGB stars \citep{den08,sok12,aka13,bal13} but all point to short-lived, explosive/ballistic mass-loss events rather than to the presence of steady jets. On the other hand, the large velocity gradients of an explosive event could in principle make the presence of maser emission difficult, as this emission requires a significant amplification path with coherent velocities along the line of sight. A possibility for its detection in I15103 is that the mass loss traced by the maser emission consists of discrete parcels of gas (i.e., bullets) ejected in an explosive event. The gas of each parcel would have the required velocity coherence to produce maser emission.

\subsection{Confirmation of the nature of IRAS 15103$-$5754 as a planetary nebula}

 It is often difficult to ascertain whether an object is a PN. Such a
 classification requires different observations to discard that it
 could be a YSO or an evolved star in a phase previous to
 photoionization (AGB or post-AGB), since all these types of objects share many
 observational characteristics. Visible images and spectra of H${\alpha}$
 and [O{\sc iii}] emission are useful diagnostic tools. However, I15103 is not visible at
 optical wavelengths \citep{sua06}. In particular, H${\alpha}$ images
 do not show any emission toward the source position. For such a deeply
 obscured object, infrared and radio data must be used to determine its
 nature.

\begin{figure}
\epsscale{1}
\plotone{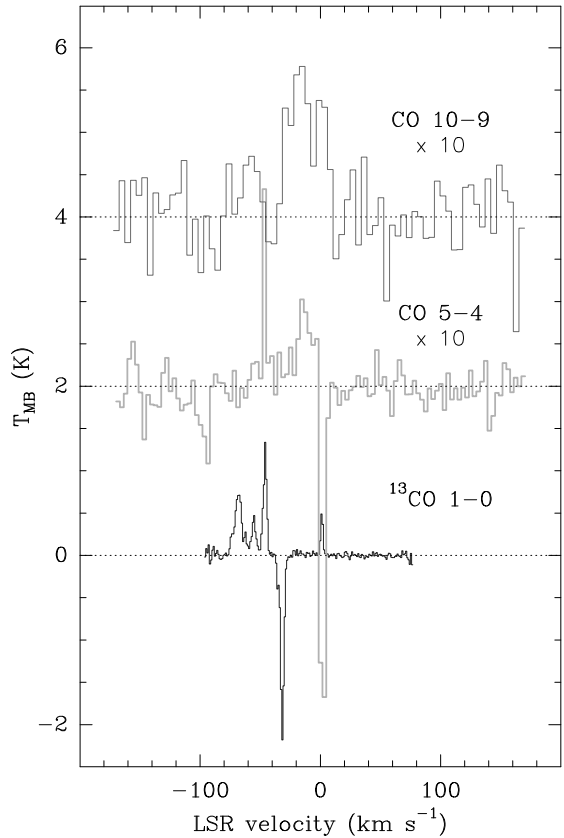}
\caption{CO spectra toward IRAS 15103$-$5754. From top to bottom: CO(J=$10\rightarrow 9$) and CO(J=$5\rightarrow 4$) obtained with Herschel (this paper), and $^{13}$CO(J=$1\rightarrow 0$) spectrum from the RMS survey \citep{urq07b}. The two Herschel spectra have been multiplied by 10 and shifted vertically, to improve their visualization\label{fig_co}}
\end{figure}

 We can immediately exclude that this object is a YSO. The position of I15103 in
 the IRAS and MSX two color diagrams is not consistent with that of an
 H{\sc ii} region \citep{sua09}. Moreover, considering its strong obscuration at optical wavelengths, if it was a YSO embedded on the parental
 cloud
 it should show significant molecular line emission associated with the
 cloud. In practice, we should detect interstellar molecular line emission with an
 LSR velocity close to that of I15103, but this is not the case.  The
 Red MSX survey (RMS) archive \citep{urq07b,lum13} shows several interstellar
 components of $^{13}$CO(J=$1\rightarrow 0$) (Fig. \ref{fig_co}). None of these components is close to that
 of the star of $\simeq -23$ km s$^{-1}$,  estimated from the maser
 position-velocity distribution (Fig. \ref{fig_image} and Section \ref{sec_maser}),  which indicates that the object is not a YSOs. However, the velocity estimate from the  maser distribution is indirect and somewhat uncertain.
 
A more direct estimate of the stellar velocity can be obtained from our Herschel data (Fig. \ref{fig_co}). The CO(J=$5\rightarrow 4$) spectrum shows some narrow emission and absorption features of interstellar origin, whose velocities coincide with those in the $^{13}$CO(J=$1\rightarrow 0$) spectrum. However,  there is an additional broader, weak component between $V_{\rm LSR}\simeq -28$ and $-8$ km s$^{-1}$, close to the central velocity estimated from the maser distribution. 
This broad component clearly dominates the CO(J=$10\rightarrow 9$) emission, which traces hotter gas than the CO(J=$5\rightarrow 4$) line (i.e., presumably closer to the central star). There could still be some interstellar contamination in the CO(J=$10\rightarrow 9$) spectrum at $V_{\rm LSR}\simeq 0$ km s$^{-1}$, but there is obvious emission around $V_{\rm LSR}\simeq -20$ km s$^{-1}$ which has no interstellar counterpart in the $^{13}$CO(J=$1\rightarrow 0$) line. Assuming that the emission around 0 km s$^{-1}$ is interstellar, the broad component would have a half-power width  $20\pm 5$ km s$^{-1}$ (centered at $-18\pm 3$ km s$^{-1}$), which is significantly broader than the interstellar CO(J=$5\rightarrow 4$) and $^{13}$CO(J=$1\rightarrow 0$) components, and is similar to the line widths observed in CO spectra of other water fountains \citep{ima09,ima12,riz13}, tracing the expansion of the circumstellar structure. We can conclude that this CO(J=$10\rightarrow 9$) emission is of circumstellar origin. The absence of significant interstellar $^{13}$CO(J=$1\rightarrow 0$) molecular line
emission associated with the source rules out its nature as a YSO.

\begin{figure}
\plotone{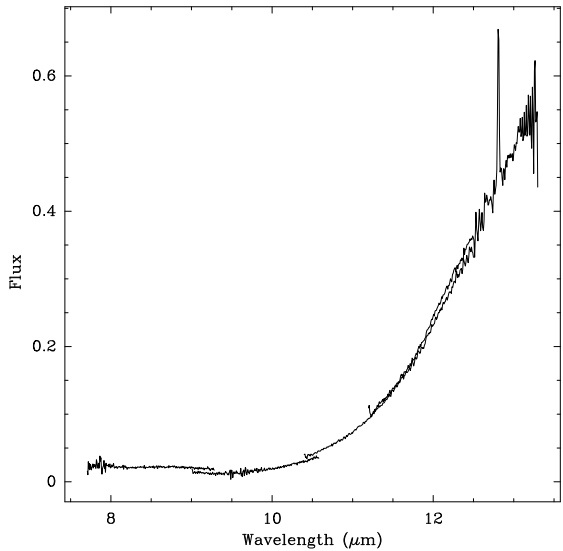}
\caption{Mid-IR spectrum toward IRAS 15103$-$5754, obtained with VISIR instrument at the VLT \label{fig_neii}. The flux is in arbitrary units. The [Ne{\sc ii}] at 12.8\,$\mu$m is clearly detected.}
\end{figure}

 Having established its evolved nature, its classification as a PN relies
 on the presence of ionized gas. This is indeed the case, since its IR
 spectrum shows a clear emission line of [Ne{\sc ii}] at 12.8\,$\mu$m
 (Fig. \ref{fig_neii}). Furthermore,  the presence of [Ne{\sc ii}]  was also hinted at the mid-IR images presented in Fig B7 of \cite{lag11}. The filter containing the [Ne{\sc ii}] line is centered at 12.81 $\mu$m and has a width of 0.42 $\mu$m, whereas the SiC filter is at a nearby wavelength (11.85 $\mu$m), but it is ten times wider (2.24 $\mu$m) and therefore, significantly more sensitive  for similar exposure times. However, note that the image in the narrower [Ne{\sc ii}] filter  seems more extended than that in the SiC one.  After rescaling  both images to get the same peak brightness, we confirmed that this extended emission in brighter in the  [Ne{\sc ii}] image. This is not expected if the emission in those filters is due to continuum alone, so the excess must be attributed to the [Ne{\sc ii}]  line emission. Those images suggest that [Ne{\sc ii}] emission is diffuse and extended, as expected in a photoionized region.
 
\subsection{Comparison with other WFs and with water-maser-emitting PNe}

There is not a well-established definition of what a WF is. The first WFs were identified based on the extremely large velocity spreads in their water maser spectra \citep[$> 100$ km s$^{-1}$][]{lik88,ima02,ima04}. However, the velocity spread itself does not constitute a reliable criterion for definition, since it depends on the relative orientation with respect to the observer. Therefore, other objects with lower velocity spreads \citep[e.g., $\simeq 50$ km s$^{-1}$][]{bob05} were later included in the WF category given their similarities with higher-velocity objects. Despite the lack of a clear-cut definition, the general trend is to consider a source as a WFs when its water maser emission traces collimated mass loss with velocities larger than that of the expansion velocities of the circumstellar envelopes in these objects \citep[10-20 km s$^{-1}$, as traced by double-peaked OH spectra with separations of 20-40 km s$^{-1}$ between the components,][]{tel89}. 

We also note that in AGB stars, the velocities of the H$_2$O maser components is always contained within the velocity range of OH maser emission. WFs show the opposite behavior, with some H$_2$O maser emission outside the velocity range of OH \citep{gom94}. However, while this is a useful observational criterion to identify new WF candidates, it does not constitute a definition of WF itself. In the case of I15103, the only published OH spectrum is that obtained with the Parkes radio telescope by \cite{tel96}, which shows a single component at $\simeq -47$ km s$^{-1}$. Therefore, it is not possible to obtain an OH velocity range. Moreover, given the large beam of the Parkes antenna at 18 cm (half power beam width $\simeq 12.6'$), there is no guarantee that the OH emission actually arises from I15103.

Therefore, if we define a WF as an object whose  water maser emission has a velocity spread  larger than that of the typical AGB envelopes, and with this emission tracing collimated mass-loss, then I15103 is a WF as far as the definition goes. However, this source shows some differential characteristics  with other WFs that are worth considering:

\begin{enumerate}
\item The water maser components in I15130 are present over the whole velocity range, without no clear clustering. The WFs IRAS 18296$-$0958 \citep{yun11} and OH 009.1-0.4 \citep{wal09} also show a similar pattern in their spectra. However, most WFs show two groups of water maser components, clearly separated in velocity \citep[e.g.,][]{ima02,bob05,gom11}, with a third central group at intermediate velocities in some objects \citep{ima13}. In the cases with clearly separated groups, the clusters at extreme velocities seem to trace the tips of a bipolar jet, and the central group could be an equatorial flow. However, in I15103 and the WFs with no clustering, the maser emission traces the mass loss along the whole body of the jet, not just the tips.
\item The ``Hubble-flow'' pattern in the position-velocity distribution of masers in I15130 (Fig. \ref{fig_image}) is not clearly seen in any other WF, which suggest that the mass-loss process in this object may be of a different nature. 
\item There is a clear bias in velocity, with most maser components being blueshifted with respect to the  velocity of the central star. The result is that only the blueshifted side of the jet is clearly collimated. There are other WFs whose maser jet and/or the spectrum is also one-sided or strongly asymmetrical, such as IRAS 18460$-$0151 \citep{ima13} or IRAS 15445$-$5449 \citep{bai09}, but most WFs are clearly bipolar. As mentioned in Sec. \ref{sec_maser}, a possible explanation for this bias is the presence of the radio continuum emission, which is amplified by the foreground (blueshifted) side of the jet by the maser mechanism. 
\end{enumerate}

All these differences with the properties of typical WFs suggest that while I15103 formally fulfil the criteria to be considered a WF, it has some unusual characteristics (although none of them are unique among WFs). 
Moreover, and what is probably more important, the discussion above also indicates that WFs show characteristics that are very different from one another, and that they do not constitute an homogeneous group. We may have to consider revising the definition of what a WF is or, alternatively, to further subdivide this class into several subcategories according to their physical characteristics. The present category of WFs may in fact include several different types of objects that only share a common characteristic (the presence of collimated, high-velocity jets traced by water masers), but that might be fundamentally different otherwise.

I15103 is also one of the few PNe showing water maser emission (hereafter H$_2$O-PNe). This category of sources comprises a total of 5 confirmed members \citep{mir01,deg04,gom08,usc14}. We note, however, that the characteristics of the maser emission in I15103 is also significantly different from the other 4 H$_2$O-PNe. The velocity spread in the water maser emission in other H$_2$O-PNe is small ($<15$ km s$^{-1}$), and most of the maser emission seem to trace equatorial tori around the central star. Only in the case of K3-35 \citep{mir01} some maser components may be related with a jet, but with a much lower velocity than in the case of WFs. Moreover, these components were transient and have not been observed thereafter \citep{deg04}.

In summary, although I15103 is formally both a WF and a H$_2$O-PN, there are significant differences with other members of these classes, and its mere inclusion in these categories should be taken with care. I15103 might be the first object of a completely different class of evolved object, whose physical characteristics might not be comparable with the rest WFs and H$_2$O-PNe. 

\subsection{Evolutionary implications of the processes observed in IRAS 15103$-$5754}

\label{sec_implications}

Even though water maser emission is present in
several PNe \citep{mir01,deg04,gom08,usc14}, the collimated high-velocity water maser emission that characterizes WFs have only been observed so far in
late AGB and post-AGB stars. 
The presence of a collimated water maser outflow in this source suggests that I15103 is an extremely young  PN, which is consistent with the results in \cite{sua14}, who {suggested that newly started photoionization is changing the spectral index of its radio continuum emission}. 

As mentioned above, this source is also unique among WFs in the sense that its water maser emission shows a linear increase of velocity with distance from the central star (a so called Hubble flow). This kinematical pattern is expected in explosive or ballistic mass-loss events. In principle, one could speculate whether maser emission in all WFs have an explosive origin, but the Hubble flow would not be evident shortly after the mass-loss event. Only as the ejected material moves away from the star, the faster gas would move farther away, and a Hubble flow would become evident. The fact that I15103 is a PN, i.e., in a later evolutionary stage than all the other known WFs, would be consistent with this time evolution of the kinematical pattern.

However, we note that the main structure of water maser emission in I15103 follows a linear distribution, with no clear gap between the maser components and the central star, whereas most WFs show their maser emission clustered on two well-separated red- and blue-shifted groups detached from the star and on each side of it \citep[e.g.][]{ima02,bob05,gom11}. If this clustered distribution were due to a single explosive event, we could expect a Hubble flow to develop with time, but the gap between the maser emission and the central star would increase, and no emission would be detected close to the star. On the other hand, a few other WFs \citep{wal09,yun11} show a more extended water maser distribution, with components close to the star, but no clear Hubble-flow pattern is present. Therefore, the mass-loss processes that give rise to water maser emission in I15103 seem to be completely different from all the other known WFs. We suggest that water maser emission in WFs that are still in the post-AGB stage is pumped mainly by jets, and this emission is seen at the bow shock created by the jet (in the cases where there are two separate clusters of water masers) or along the body of the jet (if there is a more continuous maser distribution). However, as the star enters the PN phase, high velocity water maser emission traces singular explosive events. This would indicate a fundamental change of the mass loss process in evolved stars around the end of the post-AGB phase. This type of explosive event could be a common feature at the early stages of bipolar PNe, as evidenced by the occurrence of Hubble-like flows in these objects. The Hubble-flow pattern in I15103, however, is seen at a smaller scale ($\la 850$ AU) than in other PNe and proto-PNe, and it shows a steeper velocity gradient. In particular, the Hubble flow in I15103 has a slope of $\simeq 150$ km s$^{-1}$ arcsec$^{-1}$, while it is $\simeq 2.6$  km s$^{-1}$ arcsec$^{-1}$ in the PN NGC 6302 \citep{mea08}, which is at a similar distance (1.1 kpc). In the case of an explosive event, we expect this velocity gradient to decrease with time, since the ejected material will cover a larger spatial extension. The velocity pattern we see in I15013
could represent the earliest manifestations of explosive phenomena among PNe and its precursors.

\section{Conclusions}
\label{sec_conclusions}

We presented observations of radio continuum and water maser emission at 22 GHz, as well as submm CO and mid-infrared spectra of the source IRAS 15103$-$5754, a candidate PN where single-dish observations had previously revealed a water maser spectrum that suggested the presence of a WF outflow. Our main conclusions are as follow:

\begin{itemize}
\item Water maser emission shows a wide velocity spread of $\simeq 75$ km s$^{-1}$, and it is spatially associated with the radio continuum emission. Most of the water maser components are distributed following an elongated distribution, with an orientation similar to the nebula seen in the mid-IR, and tracing a collimated high-velocity outflow. 
\item IRAS 15103$-$5754 is confirmed as a PN, based on the absence of interstellar molecular emission close to the velocity of the source (thus, ruling out it is a YSO), and the presence of ionized material. The presence of a high velocity maser outflow  indicates that the source is a WF, being the first known PN belonging to this class of objects. All previously known WFs were late AGB or post-AGB stars.
\item The velocity distribution of the maser emission shows a rough linear trend, with higher velocities farther away from the central star. Such a ``Hubble-like'' flow may be indicative of a short-lived, explosive event. This velocity pattern is not seen in other WFs, which are in previous evolutionary stages. 
\item We suggest that water maser emission in WFs that are still in the post-AGB stage is pumped mainly by  shocks in jets. However, as the star enters the PN phase, high velocity water maser emission traces singular explosive events. This would represent an important change in the mass loss processes at the end of the post-AGB phase. The ``Hubble flow'' we detected at small scales ($\la 850$ AU) could be the first manifestation of explosive events, which seem to be a common feature in the kinematics of more evolved PNe.
\end{itemize}


\acknowledgments

 JFG wishes to express
his gratitude to CSIRO Astronomy and Space Science, and the
Observatoire de la C\^ote d'Azur for their support and hospitality
during the preparation of this paper.
JFG, OS, and LFM are supported by MICINN (Spain) grant AYA2011-30228-C03-01, while MAG is suported by grant  AYA2011-29754-C03-02 (both grants include FEDER funds).  JRR acknowledges support from MICINN
grants CSD2009-00038, AYA2009-07304, and AYA2012-32032.  LFM acknowledges partial support from grant 12VI20 of the Universidad de
Vigo. LU
is supported by grant PE9-1160 of the Greek General
Secretariat for Research and Technology in the framework of the
program Support of Postdoctoral Researchers. GRL acknowledges support from
CONACyT and PROMEP (Mexico). 
This paper made use of information from the Red MSX Source survey database at http://rms.leeds.ac.uk/cgi-bin/public/RMS\_DATABASE.cgi which was constructed with support from the Science and Technology Facilities Council of the UK.

{\it Facilities:} \facility{ATCA}, \facility{Herschel}, \facility{VLT} .



\clearpage


\begin{thebibliography}{}
\bibitem[Akashi 
\& Soker(2013)]{aka13} Akashi, M., \& Soker, N.\ 2013, \mnras, 436, 1961 
\bibitem[Alcolea et 
al.(2001)]{alc01} Alcolea, J., Bujarrabal, V., S{\'a}nchez Contreras, C., Neri, R., \& Zweigle, J.\ 2001, \aap, 373, 932 
\bibitem[Bains et al.(2009)]{bai09} Bains, I., Cohen, M., 
Chapman, J.~M., Deacon, R.~M., \& Redman, M.~P.\ 2009, \mnras, 397, 1386 
\bibitem[Balick et al.(2013)]{bal13} Balick, B., 
Huarte-Espinosa, M., Frank, A., et al.\ 2013, \apj, 772, 20 
\bibitem[Boboltz 
\& Marvel(2005)]{bob05} Boboltz, D.~A., \& Marvel, K.~B.\ 2005, \apjl, 627, L45 
\bibitem[Bujarrabal et 
al.(1997)]{buj97} Bujarrabal, V., Alcolea, J., Neri, R., \& Grewing, M.\ 1997, \aap, 320, 540 
\bibitem[Bujarrabal et 
al.(2001)]{buj01} Bujarrabal, V., Castro-Carrizo, A., Alcolea, J., \& S{\'a}nchez Contreras, C.\ 2001, \aap, 377, 868 
\bibitem[Caswell 
\& Green(2011)]{cas11} Caswell, J.~L., \& Green, J.~A.\ 2011, \mnras, 411, 2059 
\bibitem[Corradi 
\& Schwarz(1993)]{cor93} Corradi, R.~L.~M., \& Schwarz, H.~E.\ 1993, \aap, 269, 462 
\bibitem[de Graauw et al.(2010)]{Gra10} de Graauw, T., Helmich, F.~P.,
Phillips, T.~G., et al.\ 2010, A\&A, 518, L6
\bibitem[de Gregorio-Monsalvo et al.(2004)]{deg04} de 
Gregorio-Monsalvo, I., G{\'o}mez, Y., Anglada, G., et al.\ 2004, \apj, 601, 
921 
\bibitem[Dennis et al.(2008)]{den08} Dennis, T.~J., 
Cunningham, A.~J., Frank, A., et al.\ 2008, \apj, 679, 1327 
\bibitem[Desmurs(2012)]{des12} Desmurs, J.-F.\ 2012, IAU 
Symposium, 287, 217 
\bibitem[G{\'o}mez et al.(2008)]{gom08} G{\'o}mez, J.~F., 
Su{\'a}rez, O., G{\'o}mez, Y., et al.\ 2008, \aj, 135, 2074 
\bibitem[G{\'o}mez et al.(2011)]{gom11} G{\'o}mez, J.~F., 
Rizzo, J.~R., Su{\'a}rez, O., et al.\ 2011, \apjl, 739, L14 
\bibitem[G{\'o}mez et al.(1994)]{gom94} G{\'o}mez, Y., 
Rodr{\'{\i}}guez, L.~F., Contreras, M.~E., 
\& Moran, J.~M.\ 1994, RMxAA, 28, 97 
\bibitem[Hrivnak et al.(2008)]{hri08} Hrivnak, B.~J., Smith, 
N., Su, K.~Y.~L., \& Sahai, R.\ 2008, \apj, 688, 327 
\bibitem[Imai(2007)]{ima07} Imai, H.\ 2007, IAU Symposium, 
242, 279 
\bibitem[Imai et al.(2002)]{ima02} Imai, H., Obara, K., 
Diamond, P.~J., Omodaka, T., \& Sasao, T.\ 2002, \nat, 417, 829 
\bibitem[Imai et 
al.(2004)]{ima04} Imai, H., Morris, M., Sahai, R., Hachisuka, K., \& Azzollini F., J.~R.\ 2004, \aap, 420, 265 
\bibitem[Imai et al.(2009)]{ima09} Imai, H., He, J.-H., 
Nakashima, J.-I., Ukita, N., Deguchi, S., 
\& Koning, N.\ 2009, \pasj, 61, 1365 
\bibitem[Imai et al.(2012)]{ima12} Imai, H., Chong, S.~N., 
He, J.-H., et al.\ 2012, \pasj, 64, 98 
\bibitem[Imai et al.(2013)]{ima13} Imai, H., Deguchi, S., 
Nakashima, J.-i., Kwok, S., \& Diamond, P.~J.\ 2013, \apj, 773, 182 
\bibitem[Izzo et al.(2004)]{izz04} Izzo, C., Kornweibel, N., 
McKay, D., et al.\ 2004, The Messenger, 117, 33 
\bibitem[Lagadec et al.(2011)]{lag11} Lagadec, E., Verhoelst, 
T., M{\'e}karnia, D., et al.\ 2011, \mnras, 417, 32 
\bibitem[Lagage et al.(2004)]{lag04} Lagage, P.~O., Pel, 
J.~W., Authier, M., et al.\ 2004, The Messenger, 117, 12 
\bibitem[Likkel 
\& Morris(1988)]{lik88} Likkel, L., \& Morris, M.\ 1988, \apj, 329, 914 
\bibitem[Lumsden et al.(2013)]{lum13} Lumsden, S.~L., Hoare, 
M.~G., Urquhart, J.~S., et al.\ 2013, \apjs, 208, 11 
\bibitem[Meaburn et al.(2005)]{mea05} Meaburn, J., L{\'o}pez, 
J.~A., Steffen, W., Graham, M.~F., \& Holloway, A.~J.\ 2005, \aj, 130, 2303 
\bibitem[Meaburn et al.(2008)]{mea08} Meaburn, J., Lloyd, M., 
Vaytet, N.~M.~H., \& L{\'o}pez, J.~A.\ 2008, \mnras, 385, 269 
\bibitem[Miranda et al.(2001)]{mir01} Miranda, L.~F., 
G{\'o}mez, Y., Anglada, G., \& Torrelles, J.~M.\ 2001, \nat, 414, 284 
\bibitem[Pilbratt et al.(2010)]{Pil10} Pilbratt, G.~L., Riedinger,
J.~R., Passvogel, T., et al.\ 2010, A\&A, 518, L1
\bibitem[Ramos-Larios et 
al.(2012)]{rl12} Ramos-Larios, G., Guerrero, M.~A., Su{\'a}rez, O., Miranda, L.~F., \& G{\'o}mez, J.~F.\ 2012, \aap, 545, A20 
\bibitem[Reid et al.(2009)]{rei09} Reid, M.~J., Menten, 
K.~M., Zheng, X.~W., et al.\ 2009, \apj, 700, 137 
\bibitem[Rizzo et 
al.(2013)]{riz13} Rizzo, J.~R., G{\'o}mez, J.~F., Miranda, L.~F., et al.\ 2013, \aap, 560, A82 
\bibitem[Roelfsema et al.(2012)]{Roe12} Roelfsema, P.~R., Helmich,
F.~P., Teyssier, D., et al.\ 2012, \aap, 537, A17
\bibitem[Sahai 
\& Trauger(1998)]{sah98} Sahai, R., \& Trauger, J.~T.\ 1998, \aj, 116, 1357 
\bibitem[Sault \& Killeen(2004)]{sau04} Sault, B., \& Killeen, N. 2004, The Miriad User Guide, http://www.atnf.csiro.au/computing/software/miriad/
\bibitem[Soker 
\& Kashi(2012)]{sok12} Soker, N., \& Kashi, A.\ 2012, \apj, 746, 100 
\bibitem[Su{\'a}rez et 
al.(2006)]{sua06} Su{\'a}rez, O., Garc{\'{\i}}a-Lario, P., Manchado, A., et al.\ 2006, \aap, 458, 173 
\bibitem[Su{\'a}rez et al.(2008)]{sua08} Su{\'a}rez, O., 
G{\'o}mez, J.~F., \& Miranda, L.~F.\ 2008, \apj, 689, 430 
\bibitem[Su{\'a}rez et 
al.(2009)]{sua09} Su{\'a}rez, O., G{\'o}mez, J.~F., Miranda, L.~F., et al.\ 2009, \aap, 505, 217 
\bibitem[Su\'arez et al.(2014)]{sua14} Su\'arez, O., G\'omez, J.~F., Bendjoya, Ph., et al.\ 2014, \apj, submitted
\bibitem[te Lintel Hekkert et 
al.(1989)]{tel89} te Lintel Hekkert, P., Versteege-Hensel, H.~A., Habing, H.~J., \& Wiertz, M.\ 1989, \aaps, 78, 399 
\bibitem[te Lintel Hekkert 
\& Chapman(1996)]{tel96} te Lintel Hekkert, P., \& Chapman, J.~M.\ 1996, \aaps, 119, 459 
\bibitem[Urquhart et 
al.(2007a)]{urq07a} Urquhart, J.~S., Busfield, A.~L., Hoare, M.~G., et al.\ 2007a, \aap, 461, 11 
\bibitem[Urquhart et 
al.(2007b)]{urq07b} Urquhart, J.~S., Busfield, A.~L., Hoare, M.~G., et al.\ 2007b, \aap, 474, 891 
\bibitem[Uscanga et 
al.(2012)]{usc12} Uscanga, L., G{\'o}mez, J.~F., Su{\'a}rez, O., \& Miranda, L.~F.\ 2012, \aap, 547, A40 
\bibitem[Uscanga et 
al.(2014)]{usc14} Uscanga, L., G{\'o}mez, J.~F., Miranda, L.~F., et al.\ 2014, MNRAS, 444, 217
\bibitem[van de Steene 
\& Pottasch(1993)]{van93} van de Steene, G.~C.~M., \& Pottasch, S.~R.\ 1993, \aap, 274, 895 
\bibitem[Vlemmings et 
al.(2014)]{vle14} Vlemmings, W.~H.~T., Amiri, N., van Langevelde, H.~J., \& Tafoya, D.\ 2014, \aap, 569, AA92 
\bibitem[Walsh et al.(2009)]{wal09} Walsh, A.~J., Breen, 
S.~L., Bains, I., \& Vlemmings, W.~H.~T.\ 2009, \mnras, 394, L70 
\bibitem[Yung et al.(2011)]{yun11} Yung, B.~H.~K., Nakashima, 
J.-i., Imai, H., et al.\ 2011, \apj, 741, 94 
\end{thebibliography}
\end{document}